\documentclass[prl,twocolumn,showpacs,preprintnumbers,amsmath,amssymb,showkeys]{revtex4}

\usepackage[usenames]{color}
\usepackage{graphicx}

\begin{document}
\title{Opto-Electrical Cooling of Polar Molecules}
\author{M. Zeppenfeld}
\email{martin.zeppenfeld@mpq.mpg.de}
\author{M. Motsch}
\author{P.W.H. Pinkse}
\author{G. Rempe}
\affiliation{Max-Planck-Institut f\"ur Quantenoptik, Hans-Kopfermann-Str. 1, D-85748 Garching, Germany}
\date{\today, PREPRINT v6}

%\tableofcontents

\begin{abstract}
We present an opto-electrical cooling scheme for polar molecules based on a Sisyphus-type cooling cycle in suitably tailored electric trapping fields. Dissipation is provided by spontaneous vibrational decay in a closed level scheme found in symmetric-top rotors comprising six low-field-seeking rovibrational states. A generic trap design is presented. Suitable molecules are identified with vibrational decay rates on the order of $100$\,Hz. A simulation of the cooling process shows that the molecular temperature can be reduced from $1$\,K to $1$\,mK in approximately $10$\,s. The molecules remain electrically trapped during this time, indicating that the ultracold regime can be reached in an experimentally feasible scheme. 
\end{abstract}

% 37.10.Mn Slowing and cooling of molecules
% 33.80.-b Photon interactions with molecules
% 37.10.-x Atom, molecule, and ion cooling methods


\maketitle

%{\bf Introduction}
The ability to prepare samples of ultracold molecules opens up exciting new possibilities in physics and chemistry, including ultrahigh-precision molecular spectroscopy and interferometry~\cite{Veldhoven04,Gerlich07}, investigations of anisotropic collisions in quantum-degenerate gases~\cite{Baranov02}, steering of chemical reactions~\cite{Krems05}, tests of fundamental physics such as the search for the electron dipole moment~\cite{Hudson02}, and novel approaches to quantum computing and quantum simulations~\cite{Tesch02,DeMille02,Micheli06}. Reaching ultracold temperatures through laser cooling has the great advantage that it does not lead to particle loss and that it is a single-particle process which does not require suitable collision properties or high densities. However, laser cooling has so far only been demonstrated for atoms and ions with simple energy-level structures, whereas optical cooling of molecules has proven confoundingly difficult.

Optical cooling of molecules requires a change in paradigm: In contrast to ultracold atoms, for which efficient cooling was realized early on~\cite{MetcalfLaserCooling} but trapping proved to be a challenge due to the shallow optical and magnetic potentials available, electric trapping of polar molecules is relatively easy~\cite{Bethlem00} and has, in fact, been demonstrated for molecules without any cooling~\cite{Rieger05}. Optical cooling of molecules could therefore start with trapped molecules and exploit the tremendous ($\sim\!1$\,K) energy-level shifts producible by laboratory electric fields, circumventing the usual requirements of standard laser cooling such as highly closed transitions, fast decay rates, and significant photon momentum transfer.

Making use of the aforementioned paradigm shift we here present a cooling scheme for molecules which is conceivable with present technology. Specifically, we replace photon recoil by an electric-field interaction energy as the means to remove energy from a molecular ensemble in a configuration reminiscent of Sisyphus cooling and single-photon cooling~\cite{Pritchard83,Soeding95,Price08}. Spontaneous emission of photons serves only to remove entropy. As a result, the number of scattered photons required to achieve substantial cooling is dramatically reduced. Slowly decaying vibrationally excited states, generally offering stricter selection rules than electronic transitions, can therefore be used for the spontaneous decay.

%The difficulty in realizing a cooling scheme for molecules is often explained by the lack of
%In this paper we present a cooling scheme which is applicable to a wide range of polar molecules

%need millions of scatterings
%laser in deep UV
%bad selection rules for electronic transitions
%low recoil in IR
%nonexistent scattering rates for far detuned rayleigh scattering\\

%cooling scheme\\
%accumulation scheme\\
%selection rules for symmetric rotor\\
%decay rates, dipole moments for various molecules\\
%Trap design\\
%simulation of cooling rates\\

\begin{figure}[bth]
\centering
\includegraphics{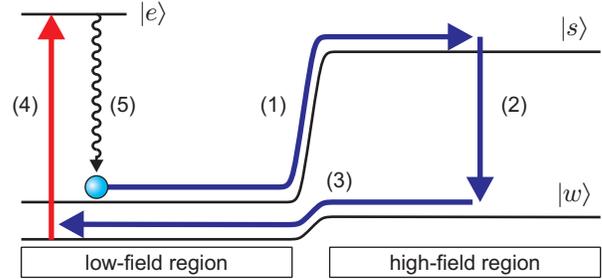}
\caption{(color online). Energy-level and state-transition diagram for the cooling scheme. A molecule in the strong lfs state $|s\rangle$ diffuses from the low-field region to the high-field region $(1)$ where it is driven to the weak lfs state $|w\rangle$ $(2)$. After moving back to the low-field region $(3)$, the molecule is driven to the excited state $|e\rangle$ $(4)$ from which it decays spontaneously to $|s\rangle$ $(5)$. The irreversible spontaneous decay makes this cycling process unidirectional.}\label{simple cooling scheme}
\end{figure}

\begin{figure}[t]
\centering
\includegraphics[width=0.5\textwidth]{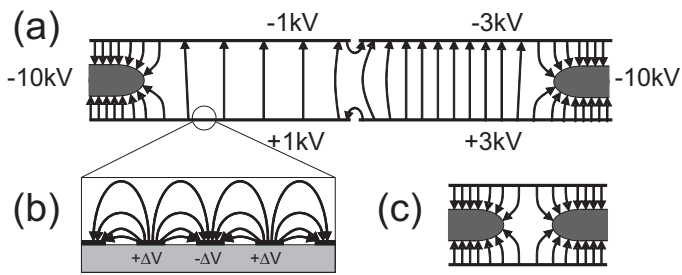}
\caption{Design of an electric trap for the cooling scheme. Regions of tunable homogeneous fields are achieved using parallel capacitor plates (a). Collisions with the plate surface are eliminated by alternatingly-charged microstructured surface electrodes~\cite{Wark92} (b). Transverse confinement is achieved by a high voltage electrode between the plates around the perimeter of the trap (a). By interrupting the perimeter electrode, an electric quadrupole guide can be connected to the trap for the injection and extraction of molecules~\cite{Rieger05} (c).}\label{Falle}
\end{figure}

%{\bf The Scheme}
Our cooling scheme is shown in Fig.~\ref{simple cooling scheme}. Two neighboring regions in space, each with a constant but different electric field, are realized by a suitable arrangement of electrodes. These electrodes also provide a high-electric-field enclosure around these regions to ensure trapping of molecules in low-field-seeking (lfs) states. Fig.~\ref{Falle} shows a possible design for the electrodes. Trapped molecules experience a potential step when moving from one region to the other. The magnitude of this potential step depends on the average orientation of the electric dipole moment of a molecule with respect to the electric field and may vary significantly for different molecular states. For one strong and one weak lfs molecular state we obtain a potential as a function of position as depicted by the curves $|s\rangle$ and $|w\rangle$ in Fig.~\ref{simple cooling scheme}.

Suppose a molecule possesses an excited state $|e\rangle$ which decays into the states $|w\rangle$ and $|s\rangle$. We induce transitions between $|w\rangle$ and $|e\rangle$ in the low-field region of the trap and transitions between $|s\rangle$ and $|w\rangle$ in the high-field region. Doing so creates a unidirectional cycling process. During the cycle, the molecule loses a kinetic energy corresponding to the difference between the potential steps of the strong and the weak lfs state, leading to overall cooling.

The main advantage of this cooling scheme is the large amount of kinetic energy which can be removed from a molecule for each spontaneously emitted photon. For a representative dipole moment $\mathbf{d}_{el}$ of $1$\,Debye [D], oriented in an electric field $\mathbf{E}$ of $100$\,kV/cm, one obtains an interaction energy of $\mathbf{d}_{el}\cdot\mathbf{E}=\frac{3}{2}k_B\times1.61$\,K. Starting with an ensemble of molecules with a translational temperature below $1$\,K, this in principle allows the removal of all of a molecule's kinetic energy in a single step. In practice, however, more than one spontaneous decay is necessary to cool a molecule: When the fields are kept constant, a molecule will generally end up in state $|s\rangle$ in the low-field region with insufficient energy to move back to the high-field region but with at least the amount of energy obtained when moving from the high- to the low-field region in the state $|w\rangle$. Further cooling to lower temperatures therefore requires the height of the electric-field step to be slowly ramped down, allowing the cooling cycle to repeat. Nonetheless, a few dozen spontaneous decays are more than enough to cool a molecule to below a mK.

Due to the small number of spontaneous photon emissions, the requirements imposed on the emission process are much less stringent than for standard laser cooling. Not only is the branching ratio for decay from the excited state to desired and undesired states much less critical, but the rate at which such transitions occur may also be much lower. As a result, the use of vibrational transitions for the spontaneous decay process is possible.

The advantage of vibrational compared to electronic excitations is that except in the case of strong resonances with other vibrational states, a molecule with one quantum of excitation in a single vibrational mode will decay primarily back to the vibrational ground state. Additionally, compared to the deep ultraviolet wavelengths required to excite electronic states of most simple chemically-stable molecules, many molecules have strong vibrational transitions in the wavelength range $3-10\,\mu$m. The coverage of this wavelength range by tunable narrow-band light sources has been significantly improved in recent years by the commercial availability of quantum-cascade lasers and optical parametric oscillators, in addition to, e.g., lead-salt lasers.

\begin{table}
\centering
\begin{tabular}{l|c|c|c||l|c|c|c}
molecule&	$d_{el}$&	$f_{\rm vib}$& $\gamma$	& molecule&	$d_{el}$&	$f_{\rm vib}$& $\gamma$\\
        &	[D]&	[cm$^{-1}$]& [Hz]&	&	[D]&	[cm$^{-1}$]& [Hz]\\
\hline
CFH$_3$&		1.85&		2964&		{\it 37 } & CF$_3$Cl&			0.50&		1105&		{\it 73 }\\
\hline
CF$_3$H&		1.65&		3036&		{\it 65 } & CF$_3$Br&			0.65&		1089&		{\it 74 }\\
\hline
CH$_3$CCH&		0.78&	    3334&		{\it 87 } & CF$_3$I&			0.92&		1080&		{\it 61 }\\
\hline
CF$_3$CCH&		2.36&		3327&		{\it 79 } & BH$_3$CO&			1.80&		{\it 2217}&	{\it 274}\\
\hline
N(CH$_3$)$_3$&	0.61&	{\it 2933}&	{\it 200 }\\
\end{tabular}
\caption{An overview of symmetric-top molecules with strong parallel vibrational transitions with permanent dipole moment $d_{el}$~\cite{CRC handbook90}, transition frequency $f_{\rm vib}$ ~\cite{nist webbook}, and spontaneous decay rate $\gamma$. The italicized values were obtained using the quantum-chemistry package \texttt{Gaussian} \cite{Gaussian}. We have successfully produced a cold sample of each of the molecules on the left using our quadrupole guide~\cite{Rangwala03}. Note that the large hyperfine splitting in CF$_3$Cl, CF$_3$Br, and CF$_3$I complicates the straightforward application of the present scheme to these molecules.}\label{molecule candidates}
\end{table}

Beyond the closed vibrational transition, the rotational transitions must be considered. The excited vibrational state must not only decay to a manageable number of rotational states, but each of these states must be lfs so that the molecule remains trapped. Disregarding linear molecules due to their generally weaker quadratic Stark interactions, symmetric-top molecules have the most stringent selection rules for dipole transitions. Describing the rotational states of a symmetric-top molecule by the quantum numbers for the total angular momentum $J$, the angular momentum about the molecule's symmetry axis $K$ and the angular momentum about a lab-fixed axis $M$, the selection rules for a parallel transition are $\Delta J=0,\pm1$, $\Delta K=0$ and $\Delta M=0,\pm1$~\cite{molecule theory}. Furthermore, the lowest-order Stark interaction is
$\label{stark shift}
E_{\rm Stark}=-\mathbf{E}\cdot\mathbf{d}_{el}=-|\mathbf{E}| |\mathbf{d}_{el}|\frac{KM}{J(J+1)}.
$
Observing that $J\ge0$, $|K|\le J$ and $|M|\le J$~\cite{molecule theory}, we see that an excited state with $|K|=J\ge2$ and $M=-K$ best satisfies the conditions stated above. Such a state may decay into a total of only five rotational states, all of which are lfs. These can be repumped using additional lasers or microwave fields. Note that the condition of few decay channels to purely lfs states can also be satisfied for linear molecules, e.g., in a $\Sigma$ electronic state using a vibrationally excited state with $M=0$, $J\ge3$~\cite{molecule theory}.

\begin{figure}[t]
\centering
\includegraphics{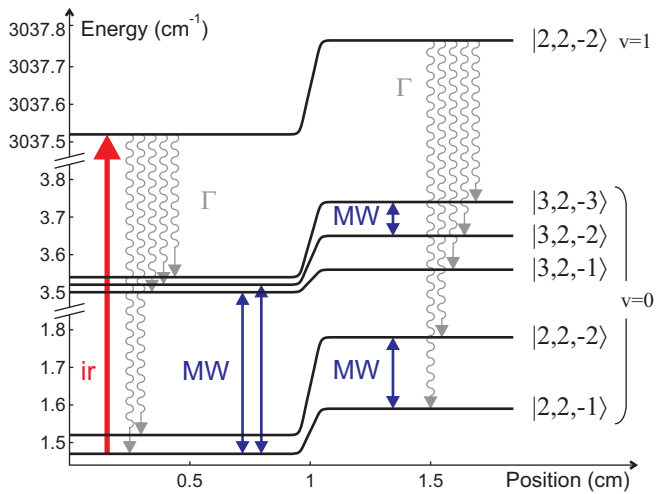}
\caption{(color online). Transition scheme used to simulate cooling of CF$_3$H. Rotational states are labeled using the notation $|J,K,M\rangle$. MW and ir denotes the induced microwave and infrared transitions. $\Gamma$ denotes spontaneous decay to the five states with vibrational excitation v$=0$. The energy levels are obtained by diagonalizing the rigid-rotor Hamiltonian for CF$_3$H at field strengths of $5$\,kV/cm and $20$\,kV/cm for the low-field and high-field region, respectively.}\label{full cooling scheme}
\end{figure} 
Use of vibrational excitations for opto-electrical cooling requires molecules with a sufficiently fast vibrational spontaneous decay rate. Table~\ref{molecule candidates} lists promising symmetric-top molecules. Although a decay rate of $\sim\!100$\,Hz is glacial relative to decay rates used for laser cooling of atoms, it is adequate considering the small number of decays needed for the scheme presented here. Nonetheless, the spontaneous decay rate raises the question how fast opto-electrical cooling progresses. This is studied by numerically solving rate equations for cooling of CF$_3$H. The rate equations and their derivation are included in the appendix. The first excitation of the C-H stretch mode at $3036$\,cm$^{-1}$ in the rotational state $J\!=\!K\!=\!-M\!=\!2$ is used as the excited state. The fact that this state spontaneously decays to five v$=0$ states (v being the vibrational quantum number) necessitates a somewhat more complicated transition scheme than the one shown in Fig.~\ref{simple cooling scheme}. Specifically, we simulate cooling using the transition scheme shown in Fig.~\ref{full cooling scheme}. The IR transition as well as each of the microwave transitions are driven with a rate of $10$\,kHz. Assuming a Stark-broadening of $10$\,MHz, this would require narrow-linewidth sources with an intensity on the order of $1$\,mW/cm$^2$ for all the transitions involved. Spontaneous decay from the excited state is modeled using a rate of $65.2$\,Hz, partitioned among the states with v$=0$ based on rigid-rotor dipole-transition matrix elements~\cite{molecule theory}. The volumes of both trap regions are set to $100$\,mm$^3$, connected by an area of $10$\,mm$^2$.

At time $t=0$, molecules are distributed among the states v$=0$ in both trap regions with a $v^2\,dv$ velocity distribution up to a cut-off velocity of $11.7$\,m/s. This is the maximal trappable velocity of the involved states due to higher-order Stark shifts. The electric-field-strength difference between the two trap regions is a free parameter which is varied as a function of time to be proportional to the $80$th percentile of the kinetic energy of the molecules. The potential-energy step for each of the molecular states is modeled using the first-order Stark shift $E_{\rm Stark}$.

\begin{figure}[t]
\centering
\includegraphics[width=0.48\textwidth]{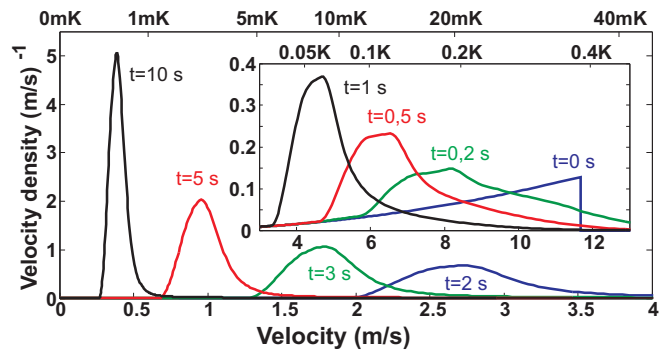}
\caption{(color online). Velocity distribution in the weak-field region of the trap after cooling for a time $t$. Velocities are converted to temperatures according to $\frac{m}{2}v^2=\frac{3}{2}k_BT$. Monitoring the population in the excited state during the cooling process shows that on average a molecule spontaneously decays $4.7\times$ during the first second, $17.0\times$ during the first $5$ seconds, and $9.0\times$ during the next $5$ seconds.%{\color{Gray}For large $t$ almost half the spontaneous decays originate from the $\sim\!10\%$ of molecules which are no longer cooled efficiently due to their energy being much larger than the potential step between the two trap regions.}
}\label{cooling}
\end{figure}

The rate of the cooling process is influenced by three effects. Most significantly, the rate coefficients indicate the time in which $1-1/e$ of molecules perform some process, whereas the time in which $99\%$ of molecules perform this process takes significantly longer. Ramping down the electric-field step too rapidly therefore causes the final energy of most molecules to substantially exceed the field-step energy so that efficient cooling is no longer possible. Secondly, the fraction of energy removed during each cooling cycle is below unity. Reducing the temperature by, e.g., a factor of ten requires several cooling cycles. Finally, spontaneous decay to the states $|2,2,-1\rangle$, $|3,2,-1\rangle$, and $|3,2,-2\rangle$ in the low-field region of the trap has no net effect, reducing the effective decay rate.

The velocity distribution of the molecules in the low-field region of the trap for various times after cooling commences is shown in Fig.~\ref{cooling}. As can be seen, significant cooling occurs in under a second. Note that the cooling rate decreases significantly as time progresses. For high temperatures, the cooling rate is limited by the decay rate of the vibrationally excited state, allowing the temperature to decrease exponentially with time. For low temperatures, the cooling rate is limited by the time it takes for the molecules to move between the two regions of the trap, with the cooling rate proportional to the velocity of the molecules. Therefore, at very low temperature the cooling process is no longer efficient, and the molecules must either be moved to a smaller trap or a different cooling scheme must be applied.

The elementary description of opto-electrical cooling so far glosses over several issues which must be addressed to ensure the experimental viability of the method. In particular, achieving required trapping times, sufficient mixing of the individual velocity components, and validity of approximate selection rules are now discussed.

In addition to collisions with the background gas, Majorana flips and rovibrational heating by thermal blackbody radiation are the identified loss channels for polar molecules stored in electric traps~\cite{Kirste09,Hoekstra07}. Although rotational heating is a problem for extremely light molecules~\cite{Hoekstra07} and vibrational heating for heavy molecules, neither is the case for the molecules considered in table~\ref{molecule candidates}. For example, the heating rates to the lowest vibrational modes never exceed a few mHz at $300$\,K for CF$_3$CCH, the heaviest molecule in table~\ref{molecule candidates}.

Majorana flips are expected to have been a problem in past trap designs with a near-zero electric field in the central trap region~\cite{Kirste09}. However, the trap in Fig.~\ref{Falle} is specifically designed to allow a homogeneous offset field throughout the vast majority of the trap volume. Furthermore, field zeros near the edges of the trap can be reduced to singular points through clever electrode design, which essentially eliminates Majorana flips.

%These have been shown to allow trap lifetimes of several seconds in past experiments~[?], enough to allow opto-electrical cooling to set in. Furthermore, peculiarities of the electric trap design outlined in Fig.~\ref{Falle} as well as of the molecule species recommended for opto-electrical cooling in table~\ref{molecule candidates} have the potential to suppress Majorana flips and rovibrational heating, respectively, such that significantly longer trapping times become possible. The trap in Fig.~\ref{Falle} is specifically designed to allow a homogeneous offset field throughout the vast majority of the trap volume. Through clever electrode design, field zeros near the edges of the trap can be reduced to individual singular points, which should essentially eliminate Majorana flips. Rotational heating by blackbody radiation is only significant for extremely light molecules, generally those with a single non-hydrogen/non-deuterium nuclei, due to the $\omega^2\,d\omega$ dependence of the phase space density of the blackbody radiation. Conversely vibrational heating is mainly a problem for heavier molecules where lowest vibrational modes are not frozen out. Even for the heaviest molecule in table~\ref{molecule candidates}, CF$_3$CCH, the heating rates to the lowest vibrational modes never exceed a few mHz at $300$\,K and are completely negligible at $77$\,K.

Opto-electrical cooling only removes energy from a single component of the velocity vector, making sufficient mixing of the velocity components a necessity. Electric-field inhomogeneities near the microstructured plate surface allow such mixing on a sufficiently short timescale. This is demonstrated by trajectory simulations discussed in the appendix.

The zero-field rigid-rotor harmonic-oscillator selection rules used so far imply a closed six-level system for opto-electrical cooling. These selection rules are modified in several ways for real molecules. Transitions with $\Delta K\ne0$ and decay to other excited vibrational states are generally possible for symmetric top molecules via resonances between near-degenerate vibrationally excited states. Due to the few spontaneous emissions needed, such couplings will at most cause problems for individual molecule species.

For non-zero values of the electric field, $J$ ceases to be a good rotational quantum number and spontaneous decay with $|\Delta J|\ge2$ becomes possible. The resulting consequences were checked by diagonalizing the rigid-rotor Hamiltonian for non-zero electric fields using molecular constants of CF$_3$H and calculating dipole transition matrix elements between the new eigenstates. Although the partitioning of spontaneous decay from the state v$=1$, $|2,2,-2\rangle$ to the five states v=0, $J=2$ and 3 is significantly changed already at electric fields of $\sim50$\,kV/cm, the spontaneous decay to states with $J\ge4$ remains below $1\%$ for fields up to $100$\,kV/cm. This effect on opto-electrical cooling is therefore negligible.

%-radiation into trap

%As discussed above, we may be chose an excited state which decays spontaneously into as few as five rotational states leading to a cooling scheme slightly more complicated than the one originally shown in Fig.~\ref{simple cooling scheme}. Specifically, choosing an excited state with $J=K=-M=2$ allows a cooling scheme to be implemented as shown in Fig.~\ref{full cooling scheme}.

%using CF$_3$H
%excited state with $J=K=-M=2$
%optical transitions between rotational states of the vibronic ground state
%Fig.~\ref{full cooling scheme}
%value of $\Delta E_a$
%initial condition
%parameters given in table: decay rates, transition rates, volumes, areas, molecule mass
%results
%discussion

%{\bf Outlook}
%The starting condition used in the simulation, namely a molecule temperature of $1$\,K, is realistic as has been demonstrated in several experiments~\cite{Doyle04}. The temperature range therefore likely to be bridged by the cooling technique proposed here, from $\sim\!1$\,K to below $1$\,mK, is possibly the most difficult one on the way to even colder molecules.

Achieving a temperature below $1$\,mK through opto-electrical cooling would allow other cooling schemes, requiring longer interaction times or higher phase-space density, to be implemented. Specifically, opto-electrical cooling can easily be extended to an accumulation scheme, for example to load molecules into a tightly confining optical dipole trap. The low temperatures and high densities thus achieved create extremely favorable starting conditions for a number of further cooling schemes such as evaporative cooling, cavity cooling, or sympathetic cooling with ultracold atoms.

\begin{acknowledgments}
Support by the Deutsche Forschungsgemeinschaft via the excellence cluster "Munich Centre for Advanced Photonics" and via EuroQUAM (Cavity-Mediated Molecular Cooling) is acknowledged.
\end{acknowledgments}

\bibliographystyle{unsrt}

\vspace{1cm}{\color{White}end}
\pagebreak
\appendix
\section{Derivation of the Rate Equations}
The rate equations used to simulate cooling of CF$_3$H can be derived as follows. The ensemble of molecules in the trap is represented by the number of molecules $p_a^{(i)}(v,t)dv$ with velocity between $v$ and $v+dv$ in the molecular state $a$ in part $i=1,2$ of the trap. Here, $i=1$ denotes the low-field region and $i=2$ denotes the high-field region of the trap.

In this description of the molecular ensemble, we ignore the position of the molecules within each trap region as well as the direction of the velocity vector $\mathbf{v}$ with $v=|\mathbf{v}|$, effectively assuming instantaneous spatial redistribution of the molecules within each trap region as well as instantaneous redistribution of the direction of $\mathbf{v}$. Whereas the assumed instantaneous spatial redistribution should at most slightly affect the validity of the results of the simulation since the molecule's thermal motion will rapidly redistribute the molecules through the trap, the assumed mixing of the components of $\mathbf{v}$ is less obvious. In fact, in the case of a potential energy in the trap which is completely separable in Cartesian coordinates, no mixing of the velocity components would occur at all.

\begin{figure}[t]
\centering
\includegraphics[width=0.48\textwidth]{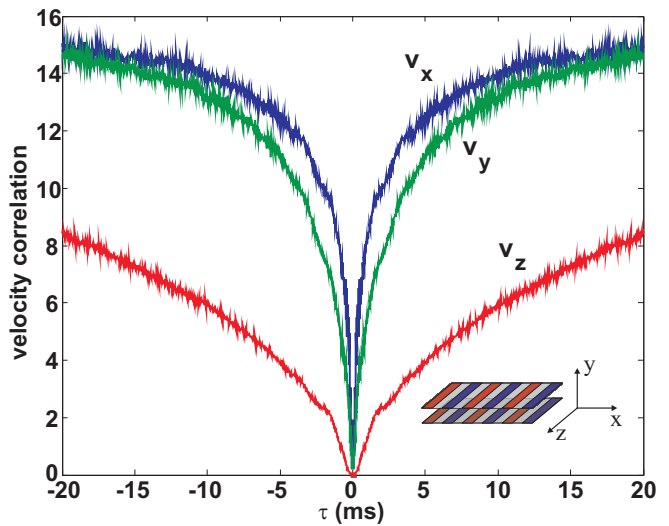}
\caption{Temporal correlation $\langle(|v_i(t)|-|v_i(t+\tau)|)^2\rangle$ of the individual components of the velocity for a particle in a trap based on the design in Fig.~\ref{Falle} of the paper. The inset shows the orientation of the velocity components relative to the trap.}\label{correlation}
\end{figure}

To address this question, molecule trajectory simulations in a trap based on the design in Fig.~\ref{Falle} of the paper were performed and the temporal correlation of the magnitude of the individual components of $\mathbf{v}$ was calculated. The calculated correlation for a particle with a velocity of $10$\,m/s in the homogeneous-field region of the trap is shown in Fig.~\ref{correlation}. Particularly for large $\tau$, the correlation function for $v_x$ and $v_y$ can be accurately reproduced by assuming a $\sim20$\,\% probability for the velocities to completely mix for each collision with the microstructured plate surface. This demonstrates that significant mixing of the velocity components occurs due to the field inhomogeneities near the plate surfaces. Due to the translational symmetry of the microstructures along the $z$-direction, this mixing does not include the $z$-component of the velocity, leading to significantly slower mixing for this velocity component. However, by arranging the microstructure plates such that the structure on the top and bottom plate is rotated by $90\,^{\circ}$ with respect to the other, all three velocity components mix on the shorter timescale. The resulting mixing is sufficiently strong that a major impact on the cooling rate does not occur.

To derive the rate equations for the molecule distributions $p_a^{(i)}(v,t)$, two processes must be taken into account. First, transitions between the various internal molecular states may occur in both regions of the trap. This is modeled by a fixed fraction of molecules in a given state switching to a different state per unit time interval,
\begin{equation}\label{transition rate}
\left.\frac{d}{dt}\right|_{\rm trans}p_a^{(i)}(v,t)=\sum_{a'=1}^6(c_{a',a}^{(i)}p_{a'}^{(i)}(v,t)-c_{a,a'}^{(i)}p_a^{(i)}(v,t)),
\end{equation}
with appropriate rate coefficients $c_{a,a'}^{(i)}$.

Second, molecules may diffuse between the two trap regions. For those molecules in region $i$ with a velocity component $v_x$ perpendicular to the interface between the trap regions, a fraction of $v_xA/V_i$ of the molecules attempt to enter the other trap region per unit time. Here, $A$ is the surface area of the interface between the trap regions and $V_i$ is the volume of trap region $i$. For those molecules in region 2 of the trap, all molecules attempting to enter region 1 succeed, whereas for molecules in region 1, only those molecules with $v_x^2>2\Delta E_a/m$ succeed, where $\Delta E_a$ is the potential energy difference between the two trap regions for molecules in state $a$ and $m$ is the molecular mass. Finally we need to average over the possible values of $v_x$. For an isotropic velocity distribution in three dimensions, a single velocity component is evenly distributed so that the fraction of molecules leaving region 2 per unit time is
\begin{equation}\label{diffusion 2}
\frac{1}{2v}\int_0^vdv_xv_x\frac{A}{V_2}=\frac{v}{4}\frac{A}{V_2},
\end{equation}
and the fraction of molecules leaving region 1 per unit time is
\begin{equation}\label{diffusion 1}
\frac{1}{2v}\int_{\sqrt{\frac{2\Delta E_a}{m}}}^vdv_xv_x\frac{A}{V_1}=\frac{v_{a,2}^2}{4v}\frac{A}{V_1}.
\end{equation}
Here, $v_{a,2}=\sqrt{v^2-2\Delta E_a/m}$ is equal to the velocity a molecule in region 1 will have once it has reached region 2. Note that the integrals are normalized by dividing by $2v$ since $v_x$ can be both positive and negative but only molecules with (in this case) positive $v_x$ can enter the opposite trap region.

\pagebreak
\begin{widetext}
Every molecule which leaves one trap region must enter the opposite region. Nonetheless, a complication arises since the velocity of the molecules changes when they move between the regions. Molecules which enter region 1 in state $a$ with velocity in the range $v$ to $v+dv$ must have had a velocity in the range $v_{a,2}$ to $v_{a,2}+v\,dv/v_{a,2}$ in region 2. As a result, the number of molecules in region 2 which could potentially increase the number of molecules $p_a^{(1)}(v,t)\,dv$ in region 1 with velocity in the range $v$ to $v+dv$ by switching regions is equal to $p_a^{(2)}(v_{a,2},t)\,v\,dv/v_{a,2}$. Combining this with Eqs. (\ref{diffusion 2}) and (\ref{diffusion 1}) one obtains a diffusion rate
\begin{equation}\label{diffusion rate}
\left.\frac{d}{dt}\right|_{\rm diff}p_a^{(1)}(v,t)=
\left\{\begin{matrix}
\frac{v_{a,2}}{4}\frac{A}{V_2}\times\frac{v}{v_{a,2}}p_a^{(2)}(v_{a,2},t)-\frac{A}{V_1}\frac{v_{a,2}^2}{4v}p_a^{(1)}(v,t),\hspace{5mm}&v\ge\sqrt{2\Delta E_a/m}\\\\ 
0,&v<\sqrt{2\Delta E_a/m}
\end{matrix}\right..
\end{equation}
Note that the $v$ in Eq.~(\ref{diffusion 2}) is the velocity in region 2 and is therefore replaced by $v_{a,2}$ in Eq.~(\ref{diffusion rate}). For $v^2<2\Delta E_a/m$, the diffusion rate is zero due to the potential energy step.

For molecules entering region 2 we introduce $v_{a,1}=\sqrt{v^2+2\Delta E_a/m}$ and the derivation of the diffusion rate is entirely analogous. One obtains as the final result,
\begin{equation}\begin{split}\label{rate equations}
\frac{d}{dt}p_a^{(1)}(v,t)&=\left\{\begin{matrix}\frac{A}{V_2}\frac{v}{4}p_a^{(2)}(v_{a,2},t)-\frac{A}{V_1}\frac{v_{a,2}^2}{4v}p_a^{(1)}(v,t)+\sum_{a'=1}^6(c_{a',a}^{(1)}p_{a'}^{(1)}(v,t)-c_{a,a'}^{(1)}p_a^{(1)}(v,t)),\hspace{5mm}&v\ge\sqrt{2\Delta E_a/m}\\\\
\sum_{a'=1}^6(c_{a',a}^{(1)}p_{a'}^{(1)}(v,t)-c_{a,a'}^{(1)}p_a^{(1)}(v,t)),&v<\sqrt{2\Delta E_a/m}
\end{matrix}\right.,\\
\frac{d}{dt}p_a^{(2)}(v,t)&=\frac{A}{V_1}\frac{v^3}{4v_{a,1}^2}p_a^{(1)}(v_{a,1},t)-\frac{A}{V_2}\frac{v}{4}p_a^{(2)}(v,t)+\sum_{a'=1}^6(c_{a',a}^{(2)}p_{a'}^{(2)}(v,t)-c_{a,a'}^{(2)}p_a^{(2)}(v,t)).\\
\end{split}\end{equation}
%$$\frac{d}{dt}p_a^{(1)}(v,t)=()p_a^{(2)}\left(\sqrt{v^2-\frac{2\Delta E_a}{m}},t\right)-\frac{1}{2v}\frac{A}{V_1}\left(\frac{v^2}{2}-\frac{\Delta E_a}{m}\right)p_a^{(1)}(v,t)+\sum_{a'=1}^6(c_{a',a}^{(1)}p_a'^{(1)}(v,t)-c_{a,a'}^{(1)}p_a^{(1)}(v,t))$$
%$$\frac{d}{dt}p_a^{(2)}(v,t)=()p_a^{(1)}(\sqrt{v^2+2dE/m},t)-\frac{A}{V_2}\frac{v}{4}p_a^{(2)}(v,t)+\sum_{a'=1}^6(c_{a',a}^{(2)}p_a'^{(2)}(v,t)-c_{a,a'}^{(2)}p_a^{(2)}(v,t))$$
These rate equations are used for the cooling simulations in the main text.
\color{White}
\end{widetext}

\end{document}